\documentclass[aps,twocolumn,nofootinbib,showpacs,amsmath,superscriptaddress,prl,reprint,preprintnumbers]{revtex4-1}
\usepackage{graphicx}
\usepackage{epstopdf}
\usepackage{bm}
\usepackage{epsfig}
\usepackage{graphics}
\usepackage{xspace}

\def\OMIT#1{}

\newcommand{\df}{\rm d}

\newcommand{\bea}{\begin{eqnarray}}
\newcommand{\eea}{\end{eqnarray}}

\newcommand{\gsim}{\mathrel{\rlap{\lower4pt\hbox{\hskip1pt$\sim$}}\raise1pt\hbox{$>$}}}

\newcommand{\Pythia}{\textsc{Pythia}\xspace}

\newcommand{\be}{\begin{equation}}
\newcommand{\ee}{\end{equation}}

\begin{document}
\title{Top Quark Mass Calibration for Monte Carlo Event Generators}

\author{Mathias Butenschoen}
\affiliation{II. Institut f\"ur Theoretische Physik, Universit\"at Hamburg, Luruper Chaussee 149, D-22761 Hamburg, Germany}

\author{Bahman Dehnadi}
\affiliation{University of Vienna, Faculty of Physics, Boltzmanngasse 5, A-1090 Wien, Austria}

\author{Andr\'e H. Hoang}
\affiliation{University of Vienna, Faculty of Physics, Boltzmanngasse 5, A-1090 Wien, Austria}
\affiliation{Erwin Schr\"odinger International Institute for Mathematical Physics, University of Vienna, Boltzmanngasse 9, A-1090 Wien, Austria}

\author{\\ Vicent Mateu}
\affiliation{Departamento de F\'\i sica Te\'orica and Instituto de F\'\i sica Te\'orica, IFT-UAM/CSIC,
Universidad Aut\'onoma de Madrid, Cantoblanco, 28049, Madrid, Spain}

\author{Moritz Preisser}
\affiliation{University of Vienna, Faculty of Physics, Boltzmanngasse 5, A-1090 Wien, Austria}

 \author{Iain W. Stewart\vspace{0.2cm}}
\affiliation{Center for Theoretical Physics, Massachusetts Institute of Technology, Cambridge, MA 02139, USA}

\begin{abstract}
The most precise top quark mass measurements use kinematic reconstruction methods, determining the top mass parameter of a Monte Carlo event
generator, $m_t^{\rm MC}$. Due to hadronization and parton shower dynamics, relating $m_t^{\rm MC}$ to a field theory mass is difficult.
We present a calibration procedure to determine this relation using hadron level QCD predictions for observables with kinematic mass sensitivity.
Fitting $e^+e^-$ 2-Jettiness calculations at NLL/NNLL order to \Pythia~8.205, 
$m_t^{\rm MC}$ differs from the pole mass by $900$/$600$\,MeV, and agrees with the MSR mass within uncertainties, $m_t^{\rm MC}\simeq m_{t,1\,{\rm GeV}}^{\rm MSR}$.

\end{abstract}


\preprint{IFT-UAM/CSIC-16-069, FTUAM-16-28, MIT-CTP 4820, UWTHPH 2016-13, DESY 16-150}

\maketitle

Making more precise measurements of Standard Model parameters is a major aim of the collider physics program.
The determination of the top quark mass is important due to its influence on many quantitative and conceptual  aspects for the
Standard  Model and beyond. The most precise determinations to date include  the combined result from the Tevatron~$m_t=174.34(64)$\,GeV~\cite{Tevatron:2014cka}, 
CMS Run-I~$m_t = 172.44(49)\,$GeV~\cite{Khachatryan:2015hba}, and
ATLAS Run-I~$m_t = 172.84(70)\,$GeV~\cite{Aaboud:2016igd}.

The highest precision measurements are based on direct reconstruction methods exploiting kinematic properties related to the
top quark mass, and are based on multivariate fits that depend on a maximum amount of information on the top decay final states.
This includes template and matrix element fits for distributions such as the measured invariant mass. These observables are
highly differential, depending on experimental cuts and jet dynamics.  Multipurpose Monte Carlo (MC) event generators are employed to do
the analysis, and the results are influenced by both perturbative and non-perturbative QCD effects. Thus the measured mass is the top mass
parameter $m_t^{\rm MC}$ contained in the particular MC event generator. Its interpretation may also depend in part on the MC tuning and
the observables used in the analysis. 

The systematic uncertainties from MC modeling are a dominant uncertainty in
the above measurements, but do not address how $m_t^{\rm MC}$
is related to a mass parameter defined precisely in quantum field theory that can be globally used for higher order predictions. The
relation is nontrivial because it requires an understanding of the interplay between the partonic components of the MC generator (hard
matrix elements and parton shower) and the hadronization model. In the context of top quark mass determinations it is often assumed that
MC generators should be considered as models whose partonic components and hadronization models are, through the tuning procedure, capable
of describing  experimental data to a precision that is higher than that of their partonic input.

In the past $m_t^{\rm MC}$ has been frequently identified with the pole mass. This is compatible with parton-shower implementations for massive quarks, but a direct identification is disfavored because of sensitivity to non-perturbative effects from below the MC shower cutoff $\Lambda_c\sim 1\,{\rm GeV}$. Also, the pole mass has an ${\cal O}(\Lambda_{\rm QCD})$ renormalon ambiguity, while $m_t^{\rm MC}$ does not (since partonic information is not employed below $\Lambda_c$). It has been argued~\cite{Hoang:2008xm,Hoang:2014oea} that $m_t^{\rm MC}$ has a closer relation to the MSR mass $m_t^{\rm MSR}(R\approx \Lambda_c)$, where the scale $R$ defining this scheme is close to $\Lambda_c$. The MSR mass $m_t^{\rm MSR}(R)$~\cite{Hoang:2008yj} applies the pole mass  subtraction for momentum fluctuations from above $R$ and also does not suffer from the renormalon ambiguity. 

For a given MC generator, $m_t^{\rm MC}$ can be calibrated into a field theory mass scheme through a fit of MC predictions to {\it hadron level} QCD computations for observables closely related to the distributions that enter the
experimental analyses. In this letter we provide a precise quantitative study on the interpretation of $m_t^{\rm MC}$ in terms
of the MSR and pole mass schemes based on a hadron level prediction for the variable $\tau_2$ for the production of a boosted
top-antitop quark pair in $e^+e^-$ annihilation. It is defined as:
\begin{equation}
\label{tau2def}
\tau_2= 1-\max_{{\vec n}_t}\frac{\sum_i|{\vec n}_t\cdot \vec p_i|}{Q}\,,
\end{equation}
where the sum is over the 3-momenta of all final state particles, the maximum defines the thrust axis $\vec n_{\rm t}$ and $Q$ is the
center of mass energy. In Ref.~\cite{Fleming:2007qr,Fleming:2007xt} a factorization theorem has been proven for boosted top quarks,
yielding hadron level predictions for $\tau_2$, which we refer to as 2-Jettiness~\cite{Stewart:2010tn}. For unstable top quarks it is
very close to thrust which has the sum of the 3-momenta magnitudes for final states instead of $Q$ in Eq.~(\ref{tau2def}). The $\tau_2$
distribution has a distinguished peak very
sensitive to the top mass, and is a delta function at $\tau_2^{\rm min}(m_t)=1\,-\,\sqrt{1-4 m_t^2/Q^2}$ at tree level. The peak region is dominated by dijet events where the top quarks decay inside narrow back-to-back cones and $\tau_2$ is directly related to the sum of the squared invariant masses $M^2_{a,b}$ in the two hemispheres defined
by the thrust axis $\vec n_t$, $(\tau_2)_{\rm peak }\approx (M_a^2+M_b^2)/Q^2$. Thus $\tau_2$ in the peak region is an observable with kinematic top mass sensitivity, just like those that enter the top quark mass reconstruction methods. Thus the results of our calibration study should provide information relevant for the interpretation of these measurements.

{\bf 2-Jettiness Distribution:}
The $\tau_2$ distribution in the peak region for boosted top quarks has the basic form
\begin{equation}
\frac{\df\sigma}{\df\tau_2} \!= \!\!\int\!\! \df k \bigg(\! \frac{\df\hat\sigma_{\rm s}}{\df\tau_2} +
\frac{\df\hat\sigma_{\rm ns}}{\df\tau_2} \!\bigg)\!\!\bigg(\!\tau_2-\frac{k}{Q}\!\bigg)F_{\!\tau_2}\!(k)\!
\Big[1+{\cal O} \mbox{\small$\big(\!\frac{\Lambda_{\rm QCD}}{Q},\!\frac{\Gamma_t}{m_t}\! \big)$}
 \!\Big]\!,
\end{equation}
where $\df\hat\sigma_{\rm s}/\df\tau_2$ contains the singular partonic QCD corrections
$\alpha_s^j\,[\,\ln^k(\tau_2-\tau_2^{\rm min})/(\tau_2\,-\,\tau_2^{\rm min})\,]_+$ and $\alpha_s^j\,\delta(\tau_2 \,-\, \tau_2^{\rm min})$
in the dijet limit and $\df\hat\sigma_{\rm ns}/\df\tau_2$ stands for the remaining partonic nonsingular QCD corrections. The shape function
$F_{\tau_2}$ describes the non-perturbative effects from wide-angle soft gluon radiation~\cite{Korchemsky:1999kt}. The singular partonic
contribution obeys a factorization theorem
\begin{align}\label{eq:factheo}
&\!\!\frac{\df\hat\sigma_{\rm s}}{\df\tau_2} =
H_Q^{(6)}(Q,\mu_Q)U_{H_Q}^{(6)}(Q,\mu_Q,\mu_m)H_m^{(6)}(Q,m_t,\mu_m)
  \\
&\!\!\times\!  Q\, U_{H_m}^{(5)}\Big(\frac{Q}{m_t},\mu_m,\mu_B\Big)\!\! \int \!\!{\df s}\!\!\int\!\! {\df k}\,
J_{B,\tau_2}^{(5)}\Big(\frac{s}{m_t},\mu_B,\Gamma_t,\delta m_t\Big) 
  \nonumber\\
&\!\!\times\! U_S^{(5)}(k,\mu_B,\mu_S)
\hat S_{\tau_2}^{(5)}\Big(Q[\tau_2-\tau_2^{\rm min}(m_t)] - \frac{s}{Q} - k,\mu_S\Big), 
  \nonumber
\end{align}
based on Soft-Collinear-Effective Theory~\cite{Bauer:2000ew, Bauer:2000yr, Bauer:2001ct, Bauer:2001yt}, which separates the contributions
from the hard interactions in the hard functions $H_Q$ and $H_m$, the jet function $J_{B,\tau_2}$, and the soft cross-talk between the top
and antitop jets in the partonic soft function $\hat S_{\tau}$. The jet function $J_{B,\tau_2}$ is derived in boosted HQET~\cite{Fleming:2007qr}
since the collinear top jet invariant mass in the peak region is very close to the top quark mass. It includes the collinear dynamics of the
decaying top quarks and leading top finite-width effects. The various evolution factors $U_X$ sum large logarithms.

Results for $\df\hat\sigma_s/\df\tau_2$ with next-to-leading logarithmic resummation~$+\,{\cal O}(\alpha_s)$~singular corrections
(NLL\,+\,NLO) can be found in Ref.~\cite{Fleming:2007xt}, with the addition of the virtual top quark contribution and rapidity logarithms
in $H_m$ and $U_{H_m}$ from Ref.~\cite{Hoang:2015vua}. The N$^2$LL evolution in $U_{H_Q}$ and $U_S$ is known from the massless quark case, and is consistent with the direct ${\cal O}(\alpha_s^2)$ calculation of the $J_{B,\tau_2}$ anomalous dimension~\cite{Jain:2008gb}. We implemented all the N$^2$LL order ingredients for the proper treatment of the flavor number dependence
[superscript (6) for including top as dynamic quark versus superscript (5) for excluding the top] in the RG
evolution~\cite{Gritschacher:2013pha, Pietrulewicz:2014qza}. We also include the ${\cal O}(\alpha_s)$ nonsingular corrections
$\df\hat\sigma_{\rm ns}/\df\tau_2$~\cite{BahmanPhD}.

For the shape function $F_{\tau_2}$ we use the convergent basis functions introduced in Ref.~\cite{Ligeti:2008ac} truncated to $4$
elements (where the 4-th element is already numerically irrelevant). These elements determine moments of the shape function $\Omega_i$~\cite{Abbate:2010xh,Abbate:2012jh},
which are the parameters that can also be fit together with $\alpha_s$ in event-shape analyses~\cite{Becher:2008cf,Davison:2008vx,Gehrmann:2009eh,Chien:2010kc,Abbate:2010xh,Abbate:2012jh,Gehrmann:2012sc,Hoang:2015hka}. The leading power correction $\Omega_1$ is defined in the \mbox{R-gap} scheme such that it cancels an $\mathcal{O}(\Lambda_{\rm QCD})$ renormalon
present in $\hat S_{\tau_2}$~\cite{Hoang:2007vb}. This is achieved through an appropriate subtraction series
$\delta(R_S,\mu_S)$~\cite{Hoang:2008fs} which induces both $R_S$ and $\mu_S$ dependence in $\Omega_1$. We quote results for $\Omega_1$ at
the reference scales $\mu_S=R_S=2$\,GeV. The evolution of $\Omega_1$ with $R_S$ is described by R-evolution~\cite{Hoang:2008yj,Hoang:2009yr}.

Eq.~(\ref{eq:factheo}) is written in terms of a generic mass scheme $m_t$, with $\delta m_t = m_t^{\rm pole} - m_t$ in $J_{B,\tau_2}^{(5)}$
controlling the dominant sensitivity to the mass scheme. In the pole mass scheme $\delta m_t=0$.  Using renormalon-free schemes, the
$\overline{\rm MS}$ mass with $\delta m_t\propto m_t$ is appropriate for the hard functions. In the jet function $J_{B,\tau_2}^{(5)}$ one
has to adopt a scheme such as MSR~\cite{Hoang:2008yj} with 
$\delta m_t\sim R\sim \Gamma_t$ to maintain the power
counting in the peak region. The MSR scheme is defined by ($a\equiv \alpha_s^{(5)}(R)/4\pi$)
\begin{equation}\label{eqMSR}
m_t^{\rm pole} - m_t^{\rm MSR}(R) \,\equiv \, R\,(\,c_1 a + c_2 a^2 + c_3 a^3+\ldots\,)\,,
\end{equation}
where $c_1=5.333$, $c_2=131.785$, $c_3=4699.703, \ldots$ are precisely the coefficients that define the series relating the
$\overline{\rm MS}$ to pole mass, $m_t^{\rm pole} - \overline{m}_t( \overline{m}_t)$ with $R= \overline{m}_t( \overline{m}_t)$. The
evolution of the MSR mass with $R$ is also described by R-evolution. The MSR mass is convenient as it is directly related to the
$\overline{\rm MS}$ mass, $m_t^{\rm MSR}(\overline{m}_t) = \overline{m}_t( \overline{m}_t)$. Due to
$\lim_{R\to 0}m_t^{\rm MSR}(R)=m_t^{\rm pole}$ it interpolates to the pole mass. However, in taking this limit one encounters the
Landau singularity reflecting the pole mass renormalon problem.

To sum large logarithms we use $\tau_2$-dependent scales $\mu_i(\tau_2)$ and $R_i(\tau_2)$, known as profile
functions~\cite{Ligeti:2008ac,Abbate:2010xh}. They have canonical scaling in  resummation regions, freeze at a perturbative scale to avoid
the Landau pole, and exhibit smooth transitions between regions. They are expressed in terms of $9$ parameters which are varied to estimate
perturbative uncertainties. We develop a natural generalization of those used for massless event shapes in~\cite{Hoang:2014wka}, to which they
reduce in the massless limit~\cite{BahmanPhD}.

For a given center of mass energy $Q$, the key parameters that enter the QCD factorization predictions for the $\tau_2$ distribution are
the top mass $m_t$, the top width $\Gamma_t$, the hadronic parameters $\Omega_i$, and the strong coupling $\alpha_s(m_Z)$.  We will
consider fits both in the pole and the MSR mass schemes. Our results in the MSR scheme are given in terms of
$m_t^{\rm MSR}(1\,\mbox{GeV})$ following~\cite{Hoang:2008xm,Hoang:2014oea}.

{\bf Fit Procedure:}
For a given $m_t^{\rm MC}$ we produce MC datasets for $\df\sigma/\df\tau_2$ in the peak region for various $Q$ values. For a given profile
and value of $\alpha_s(m_Z)$ we fit the parameters $m_t$ and $\Omega_i$ of the hadron level QCD predictions to this MC dataset. We fit for
integrals over bins in $\tau_2$ of size $\simeq 0.13\,{\rm GeV}/Q$. For each $Q$ value the distribution is normalized over the fit range, and multiple $Q$s are needed simultaneously to break degeneracies.
This procedure is carried out for the MC output and the QCD predictions. We then construct the $\chi^2$  using the statistical uncertainties
in the MC datasets. We do the fit by first, for a given value of $m_t$, minimizing $\chi^2$ with respect to the $\Omega_i$ parameters. The
resulting marginalized $\chi^2$ is then minimized with respect to $m_t$ used in the QCD predictions. Uncertainties obtained for the QCD 
parameters from this $\chi^2$ simply reflect the MC statistical uncertainties used to construct the $\chi^2$.  When fitting for
$m_t^{\rm pole}$ or $m_t^{\rm MSR}(1\,{\rm GeV})$ we find that the resulting $\chi^2$ is no longer sensitive to $\alpha_s(m_Z)$. Therefore
we fix $\alpha_s(m_Z)$ to the world average, and do not consider it as a fit parameter.

To estimate the perturbative uncertainty in the QCD predictions we take $500$ random points in the profile-function parameter space and
perform a fit for each of them. The $500$ sets of best-fit values provide an ensemble from which we remove the upper and lower $1.5\%$ in
the mass values to eliminate potential numerical outliers. From the ensemble we determine central values from the average of the largest
and smallest values and perturbative uncertainties from half the covered interval.

To illustrate the calibration procedure we use \Pythia~8.205~\cite{Sjostrand:2006za,Sjostrand:2014zea} with the $e^+e^-$ default tune~7
(the Monash 2013 tune~\cite{Skands:2014pea} for which $\Lambda_{c} = 0.5$\,GeV) for top mass parameter values $m_t^{\rm MC}=170$, $171$,
$172$, $173$, $174$ and $175$\,GeV. We use a fixed top quark width $\Gamma_t=1.4\,{\rm GeV}$ which is independent of $m_t^{\rm MC}$.
(Final calibration results for a $m_t^{\rm MC}$-dependent top width differ by less than $25$\,MeV). No other changes are made to the
default settings. To minimize statistical uncertainties we generate each distribution with $10^7$ events. We have carried out fits for
the following seven \mbox{Q sets} (in GeV units): $(600, 1000, 1400)$, $(700, 1000, 1400)$, $(800, 1000, 1400)$, $(600$ -- $900)$,
$(600$ -- $1400)$, $(700$ -- $1000)$ and $(700$ -- $1400)$, where the ranges refer to steps of $100$. For each one of these sets we have
considered three ranges of $\tau_2$ in the peak region: $(60\%, 80\%)$, $(70\%, 80\%)$ and $(80\%, 80\%)$, where $(x\%, y\%)$ means that
we include regions of the spectra whose $\tau_2 < \tau_2^{\rm peak}$ having cross-section values larger than $x\%$ of the peak height,
and $\tau_2 > \tau_2^{\rm peak}$ with cross sections larger than $y\%$ of the peak height, where $\tau_2^{\rm peak}$ is the peak position.
This makes a total of $21$ fit settings each of which give central values and scale uncertainties for the top mass and the $\Omega_i$.

\begin{figure}[t!]
\includegraphics[width=0.33\textwidth]{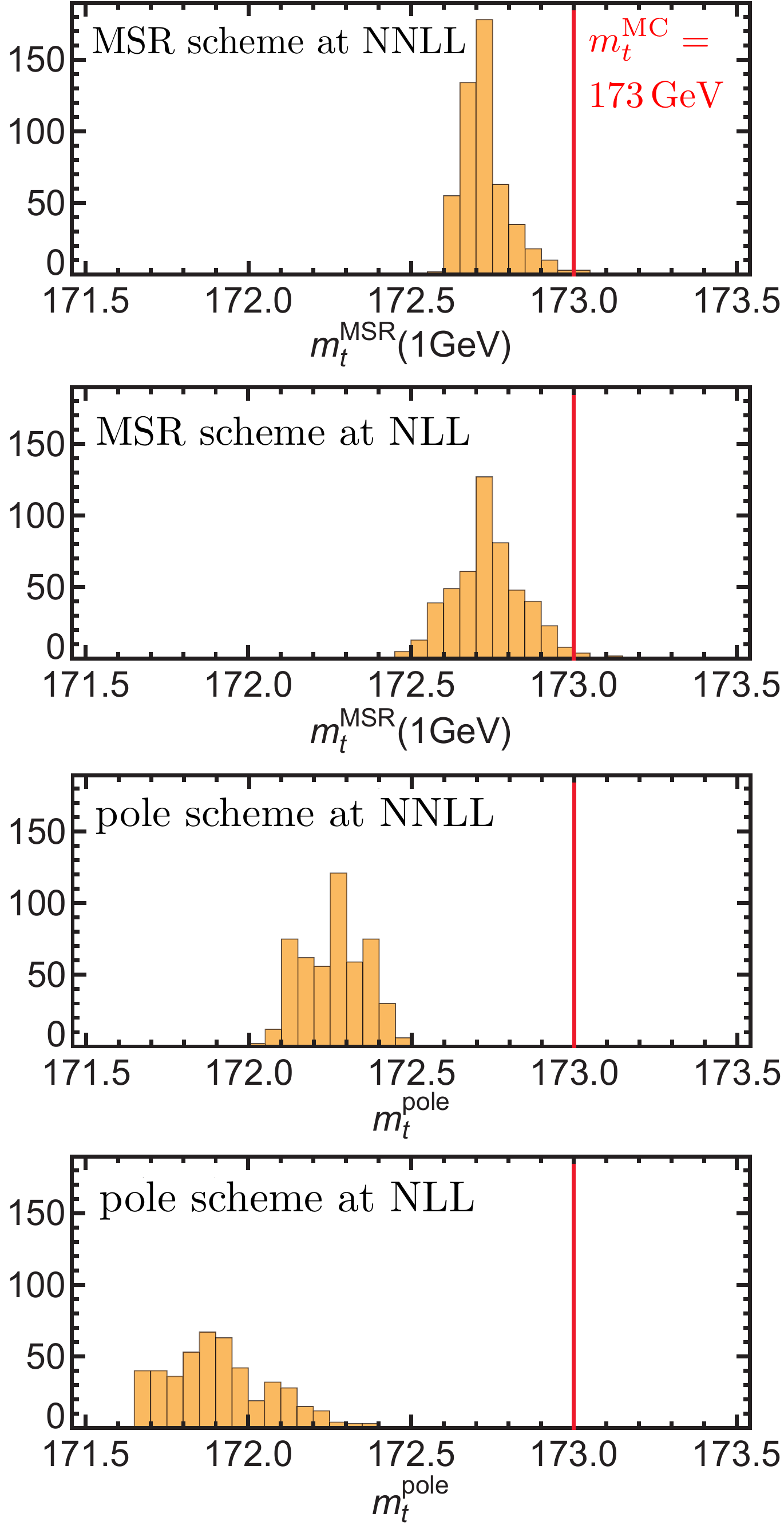}
\caption{\label{fig:histograms} 
Distribution of best-fit mass values from the scan over parameters describing perturbative uncertainties. Results are shown for cross
sections employing the MSR mass $m_t^{\rm MSR}(1\,{\rm GeV})$ (top two panels) and the pole mass $m_t^{\rm pole}$ (bottom two panels),
both at N$^2$LL  and NLL. The \Pythia datasets use $m_t^{\rm MC} = 173$\,GeV as an input (vertical red lines). }
\end{figure}

\begin{figure*}[t!]
\begin{center}
\includegraphics[width=0.32\textwidth]{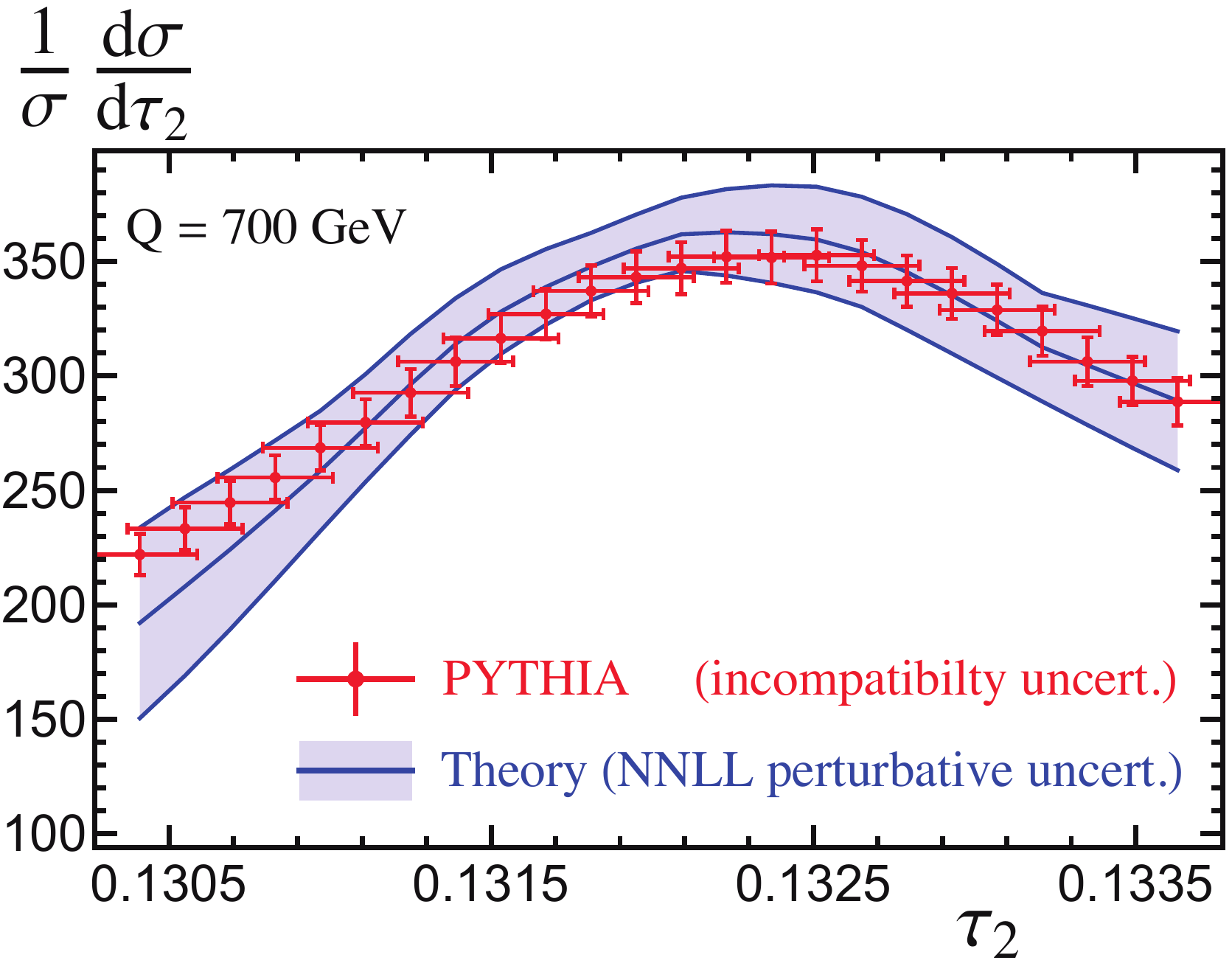}~~
\includegraphics[width=0.33\textwidth]{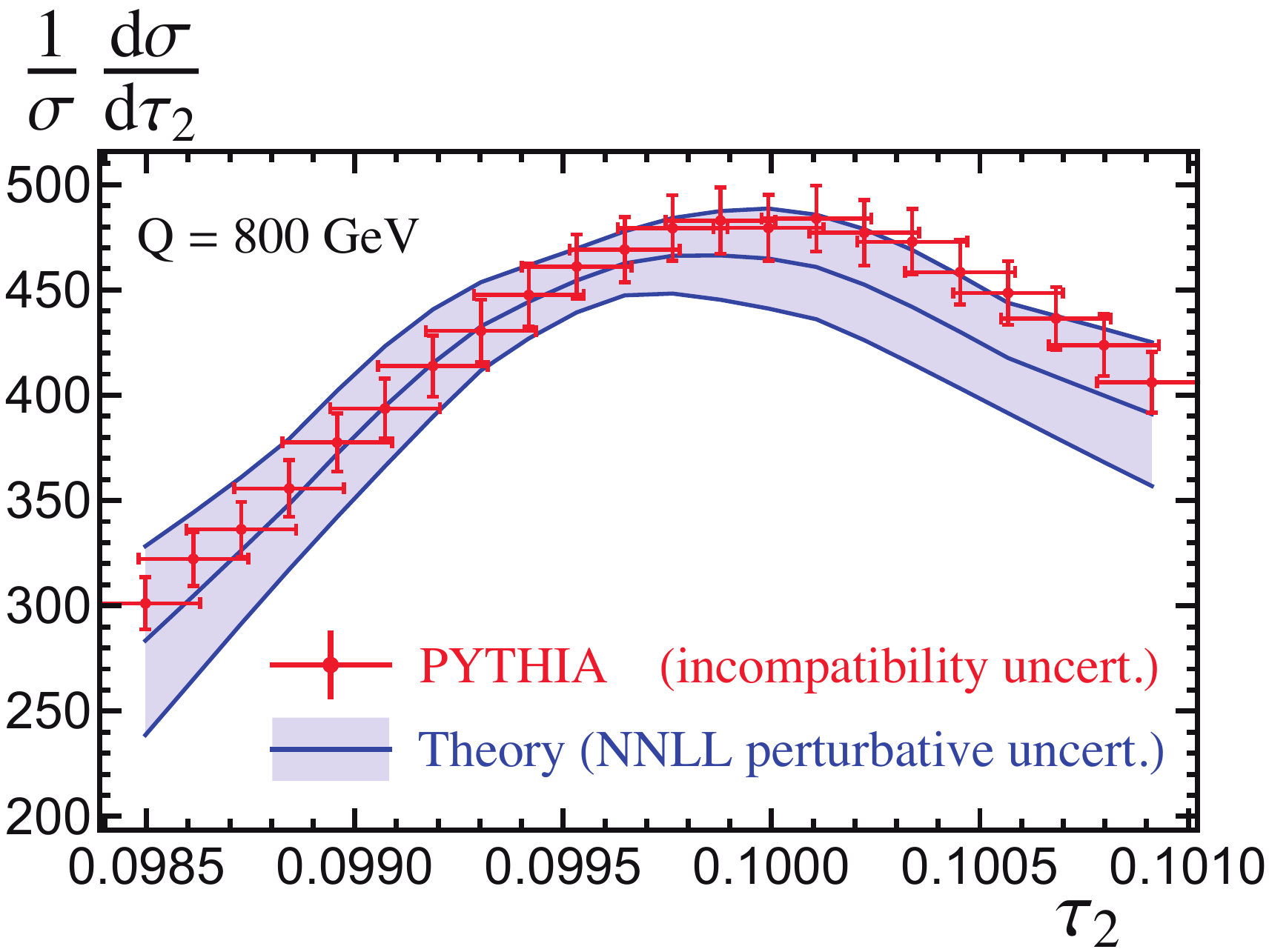}~
\includegraphics[width=0.32\textwidth]{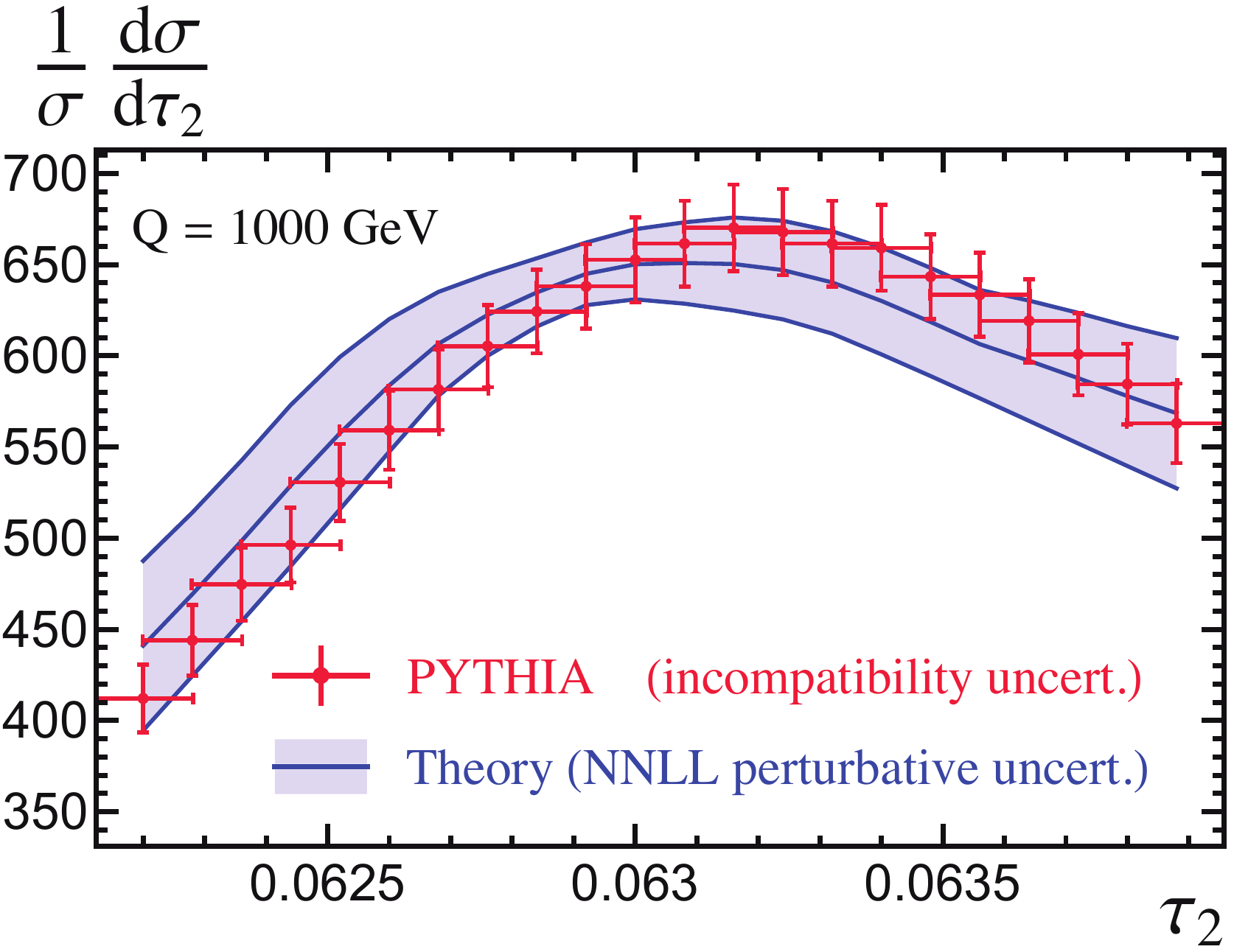}
\caption{\label{fig:data-Pythia} Comparison of \Pythia samples with $10^7$ events and $m_t^{\rm MC} = 173$~\,GeV (red dots) to the
theoretical prediction in the MSR scheme at N$^2$LL for $m_t^{\rm MSR}(1~\mbox{GeV})=172.81$\,GeV and $\Omega_1=0.44$\,GeV. The blue band
shows the perturbative uncertainty from a random scan over $500$ profile functions. Vertical error bars on the \Pythia points are obtained
by a global rescaling of \Pythia statistical uncertainties such that the average $\chi^2_{\rm min}/{\rm dof} = 1$ and roughly indicate the
incompatibility uncertainties on the cross sections. Horizontal error bars are related to the N$^2$LL incompatibility uncertainty in the
MSR mass shown in Tab.~\ref{tab:results}. }
\end{center}
\end{figure*}

{\bf Numerical Results of the Calibration:}
To visualize the stability of our fits we display in Fig.~\ref{fig:histograms} the distribution of best-fit mass values obtained for 500
random profile functions for  $m_t^{\rm MC} = 173$\,GeV based on the $Q$ set $(600-1400)$ and the bin range $(60\%, 80\%)$. Results are
shown for $m_t^{\rm MSR}(1\,\mbox{GeV})$ and $m_t^{\rm pole}$ at NLL and N$^2$LL order, exhibiting good convergence, with the higher order
result having a smaller perturbative scale uncertainty. The results for $m_t^{\rm MSR}(1\,\mbox{GeV})$ are stable and about $200$\,MeV
below $m_t^{\rm MC}$ confirming the close relation of $m_t^{\rm MSR}(1\,\mbox{GeV})$ and $m_t^{\rm MC}$ suggested in
Ref.~\cite{Hoang:2008xm,Hoang:2014oea}. We observe that $m_t^{\rm pole}$ is about $1.1$\,GeV (NLL) and $0.7$\,GeV (N$^2$LL) lower than $m_t^{\rm MC}$, demonstrating that corrections here are bigger, and {\em that the MC mass can not simply be identified with the pole mass}. These fit results are compatible with converting $m_t^{\rm MSR}$ with  $R\simeq\mu_B\simeq \mu_S Q/m_t\simeq 10$\,GeV to $m_t^{\rm pole}$ using Eq.~(\ref{eqMSR}), where $\mu_B$ is the renormalization scale of the jet function $J_{B,\tau_2}$ which governs the
dominant mass sensitivity. In Fig.~\ref{fig:data-Pythia} we see the level of agreement between the MC and theory results in the MSR scheme
at N$^2$LL order for this fit. The bands show the N$^2$LL perturbative uncertainty from the profile variations.

\begin{table}[t!]
\begin{tabular}{|llcccc|}
\multicolumn{6}{c}{\footnotesize $m_t^{\rm MC} = 173$\,GeV\ \ 
  \big($\tau_2^{e^+e^-}$\big)} \\
\hline
~mass~~~~~&order~~~& central & perturb.  &  incompatibility & ~total~\\
\hline\hline
~$m_{t,1\,\rm GeV}^{\rm MSR}$ & NLL      & $172.80$ & $0.26$ & $0.14$ & $0.29$\\
~$m_{t,1\,\rm GeV}^{\rm MSR}$ & N$^2$LL  & $172.82$ & $0.19$ & $0.11$ & $0.22$\\
~$m_t^{\rm pole}$             & NLL      & $172.10$ & $0.34$ & $0.16$ & $0.38$\\
~$m_t^{\rm pole}$             & N$^2$LL  & $172.43$ & $0.18$ & $0.22$ & $0.28$\\
\hline
\end{tabular}
\caption{\label{tab:results} Results of the calibration for $m_t^{\rm MC} = 173$\,GeV in \Pythia, combining results from all Q sets and bin
ranges. Shown are central values, perturbative and incompatibility uncertainties, and the total uncertainty, all in GeV.}
\end{table}

The results from the fits to the 21 different Q sets and bin ranges mentioned above are quite similar. The differences can be associated to
the level of incompatibility of the MC event generator results to the QCD predictions, and unlike the perturbative uncertainties these
differences need not necessarily decrease when going from NLL to N$^2$LL. We will use the differences from the 21 fits to assign an additional
{\it incompatibility uncertainty} between QCD and the MC generator for the calibration. 

To quote final results we use the following procedure:
(1) Take the average of the highest and lowest central values from the 21 sets as the final central value of our calibration.
(2) Take the average of the scale uncertainties of these sets as our final estimate for the perturbative uncertainty.
(3) Take the half of the difference of the largest and smallest central values from the sets as the incompatibility uncertainty between
QCD and the MC.
\mbox{(4) Quadratically} add the perturbative, and incompatibility errors to obtain a final uncertainty.

Using $\alpha_s$ values within the uncertainty of the world average  $\alpha_s(m_Z)=0.1181(13)$ gives an additional parametric uncertainty
of $\simeq 20$\,MeV for $m_t^{\rm MSR}(1\,\mbox{GeV})$ and $m_t^{\rm pole}$ at N$ ^2$LL order. This is an order of magnitude smaller than
the other uncertainties and we therefore neglect it.

Table~\ref{tab:results} shows our final results for the MSR mass $m_t^{\rm MSR}(1\,\mbox{GeV})$ and $m_t^{\rm pole}$ at NLL and N$^2$LL order,
utilizing the $m_t^{\rm MC}=173$\,GeV dataset. For $m_t^{\rm MSR}(1\,\mbox{GeV})$ we observe a reduction of perturbative uncertainties from
$260$\,MeV at NLL to  $190$\,MeV at N$^2$LL. The corresponding incompatibility uncertainties are $140$ and $110$\,MeV. The corresponding fit
results for the first shape function moment are $\Omega_1^{\rm PY}=0.42\pm 0.07\pm0.03$\,GeV at N$^2$LL and
$\Omega_1^{\rm PY}=0.41\pm 0.07\pm0.02$\,GeV at NLL order with the first uncertainty coming from scale variation and second from 
incompatibility. The result agrees nicely with the expectation that $\Omega_1\sim \Lambda_{\rm QCD}$. For $m_t^{\rm pole}$ there is a significant difference to $m_t^{\rm MC}$, and we observe that the central value shifts by $330$\,MeV between NLL and N$^2$LL order. There is a reduction of perturbative uncertainties like in the MSR scheme, however the 
incompatibility uncertainty increases at N$^2$LL order. 
These results may not be unexpected, since the pole mass often leads to poor convergence of perturbative series. 

\begin{figure}[t!]
\includegraphics[width=0.4\textwidth]{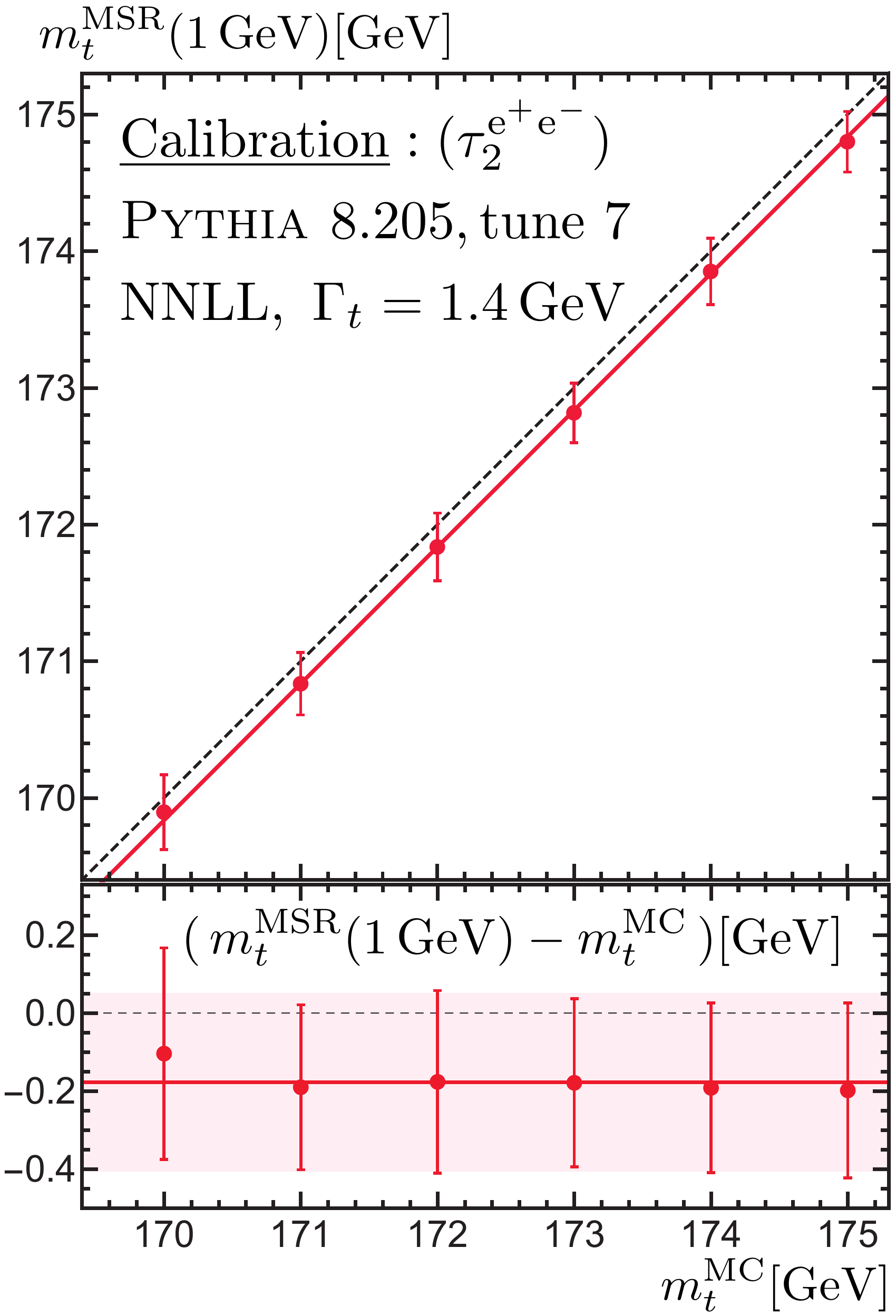}
\caption{\label{fig:moneyPlot}
Dependence of the N$^2$LL fit result for the MSR mass on the input 
$m_t^{\rm MC}$ value in \Pythia.  The error bars show the total calibration uncertainty. The red solid lines correspond to the weighted
average of the individual results. The red shaded area shows the average of the individual uncertainties. }
\end{figure}

Figure~\ref{fig:moneyPlot}, shows the outcome of our fits for the MSR mass $m_t^{\rm MSR}(1\,\mbox{GeV})$ at N$^2$LL order with six
different input values for $m_t^{\rm MC}$, and error bars with the total uncertainties.  We see the expected strong correlation between
these masses. This calibration results in Tab.~\ref{tab:results} and Fig.~\ref{fig:moneyPlot} should be independently determined for each MC and generator setting (such as different tunes).

To the extent that the treatment of the top in MC generators and QCD factorizes for different kinematically sensitive observables and from whether one considers $e^+e^-$ or $pp$ collisions, our method can be used to calibrate $m_t^{\rm MC}$ in current experimental reconstruction analyses.  $pp$ collisions introduce initial state radiation, color reconnection, and additional hadronization and multi-parton interaction effects, not present in $e^+e^-$.  In the future our method can be extended to use a $pp$ observable to directly study these effects. Prior to this, we believe that applying our $e^+e^-$ calibration to $m_t^{\rm MC}$ from a typical $pp$ reconstruction analysis will give a more accurate result than assuming $m_t^{\rm MC} = m_t^{\rm pole}$. When corresponding hadron level predictions exist, this calibration procedure can also be applied to other MC parameters.  The calibration procedure may also provide new ways to test and improve MC event generators.

{\bf Acknowledgment:}
We thank Marcel Vos, Peter Skands, Torbj\"orn Sj\"ostrand, Juan Fuster and Frank Tackmann for numerous discussions. We acknowledge partial support by the FWF Austrian Science Fund under the Doctoral Program No.~W1252-N27 and the Project No.~P28535-N27, the Spanish MINECO “Ram\'on y Cajal” program
(RYC-2014-16022), the U.S. Department of Energy under the Grant No.~\mbox{DE}-\mbox{SC0011090}, and the Simons Foundation through the Grant 327942. We
thank the Erwin-Schr\"odinger International Institute for Mathematics and Physics, the University of Vienna and Cultural Section of the
City of Vienna (MA7) for partial support.

\bibliography{thrust3}

\end{document}